\documentclass[fleqn,10pt]{wlscirep}
\usepackage[utf8]{inputenc}
\usepackage[T1]{fontenc}
\usepackage{lineno}
\usepackage[normalem]{ulem}

\title{Quantum neural networks with multi-qubit potentials}

\author[1,*]{Yue Ban}
\author[2,3]{E. Torrontegui}
\author[4,5,6]{J. Casanova}
\affil[1]{TECNALIA, Basque Research and Technology Alliance (BRTA), 48160 Derio, Spain}
\affil[2]{Departamento de F\'{\i}sica, Universidad Carlos III de Madrid, Avda. de la Universidad 30, 28911 Legan\'{e}s (Madrid), Spain}
\affil[3]{Instituto de F\'{i}sica Fundamental IFF-CSIC, Calle Serrano 113, 28006 Madrid, Spain}
\affil[4]{Department of Physical Chemistry, University of the Basque Country UPV/EHU, Apartado 644, 48080 Bilbao, Spain}
\affil[5]{EHU Quantum Center, University of the Basque Country UPV/EHU, Leioa, Spain}
\affil[6]{IKERBASQUE, Basque Foundation for Science, Plaza Euskadi 5, 48009 Bilbao, Spain}
\affil[*]{ybanxc@gmail.com}

\begin{abstract}
We propose quantum neural networks that include multi-qubit interactions in the neural potential leading to a reduction of the network depth without losing approximative power. We show that the presence of  multi-qubit potentials in the quantum perceptrons enables more efficient information processing tasks such as XOR gate implementation and prime numbers search, while it also provides a depth reduction to construct distinct entangling quantum gates like CNOT, Toffoli, and Fredkin. This simplification in the network architecture paves the way to address the connectivity challenge to scale up a quantum neural network while facilitating its training.	
\end{abstract}
\begin{document}
\flushbottom
\maketitle
%
%
\thispagestyle{empty}

\section*{Introduction}
Information is a resource due to its advance and expansion in the digitalization and control~\cite{Castells96}. However, programing explicit algorithms with good performance may become unfeasible due to the vertiginous growth in (i) the amount of available information with which classical algorithms have to deal~\cite{Walter05}, and (ii) the inherent difficulty of finding efficient algorithms for specific problems~\cite{Kolmogorov63}. All these limit our current capabilities in information processing tasks. Two alternative approaches that would contest these limitations are machine learning and quantum computing \cite{Samuel59, Feynman82}. 

On the one hand,  machine learning (ML) is a branch of artificial intelligence that uses statistical techniques to give computers the ability to progressively learn with input data without being explicitly programmed. ML is based on the generation of a hypothesis that is optimized from sample inputs and re-used to generate new predictions \cite{Hebb49}. Thus the algorithms can learn from data and overcome the static program instructions by making data-driven decisions from sample inputs. Among the distinct hypothesis models, neural networks \cite{McCulloch43} are very extended due to the blooming of deep learning \cite{Oh04, LeCun15}. Artificial neural networks are organized in layers and each layer learns new behavior patterns \cite{Kleene56}. The computational power of artificial neural networks relies on this architecture where {\it neurons} in each layer feed signals into other neurons allowing parallel-processed computing~\cite{Rosenblatt57, Hopfield84}. In this manner, several calculations can be performed at the same time, and large computational problems can often be divided into smaller ones, which can be then solved simultaneously. The versatility of neural networks to classify complex data relies on the {\it universal approximation theorem}, which leads artificial neural networks the capacity to approximate any  function \cite{Cybenko89}. As result, they span a broad range of applications such as speech \cite{Dahl12} or object recognition \cite{Hinton06}, spam filters \cite{Dada19}, vehicle control \cite{Buehler09, Devi20}, trajectory prediction \cite{Valsamis17}, decision making \cite{Kashyap18}, game-playing \cite{Furnkranz10}, or automated trading systems \cite{Huang19}. 

On the other hand, quantum computing represents a different paradigm from classical information processing. Based on an alternative information encoding that exploits the quantum properties of matter, systems that encompass several quantum bits (qubits) are exponentially hard to simulate with classical devices \cite{Feynman82} showing that quantum systems do not seem to obey the Church thesis \cite{Kleene36}, 
and consequently they are not polynomially equivalent to classical systems. Then, quantum systems harnessed as computational devices, might be dramatically more powerful than any other classical system \cite{Arute19}. The universality of quantum computing \cite{Kitaev97} expands a broad range of applications. Some illustrative examples are linear systems solvers \cite{Harrow09}, molecule simulators \cite{Peruzzo14}, combinatorial optimizers \cite{Farhi14}, black-box \cite{Grover96} and factorization problems \cite{Shor94}, or Hamiltonian simulations \cite{Lloyd96}. Although a set of single $N=1$ and two-qubit $N=2$ gates is a universal approximator \cite{Nielsen11}, larger multi-qubit ($N>2$) gates may offer a computational advantage that reduces complexity in existing algorithms \cite{Mottonen04, Reagor18, Baekkegaard19}.

Quantum machine learning \cite{Kak95, Schuld14, Biamonte17, Farhi18, Schuld19, Schuld20, Cao17, Salina20, Torrontegui19} aims for the symbiosis of both paradigms to the mutual reinforcement. To achieve improvements of the machine learning protocols by leveraging quantum resources in comparison to their classical counterparts is the goal for the future. On the other hand, the universality of artificial networks may enhance the accuracy and efficiency of quantum protocols~\cite{Carrasquilla17, Deng17, Torlai18, Aharon19, Ban20}. 
 By making analogy to classical neural networks, a quantum neural network (QNN) consists of quantum perceptrons (neurons) possessing nonlinear activation functions in different layers. In the network, the hidden layers in a QNN are the intermediate ones which are composed of quantum perceptrons, each of which is a qubit encoded in an Ising Hamiltonian \cite{Torrontegui19}. By measuring the excitation probability of the reduced eigenstate of such a Hamiltonian, one can get the nonlinear activation function.
In this work, we propose an extension of a universal QNN  enabling multi-qubit ($N>2$) interactions that lead to a reduction of the network depth while keeping the approximation power. As a result, achieving a simplification of existing protocols not only requires shorter operation times, but it may also introduce less accumulation of errors due to the reduction in the amount of requested gates. The article is structured as follows: We firstly define that the quantum perceptron is a single-output quantum neuron with multi-qubit interactions, connecting to several-input quantum neurons without hidden layers. The result of nesting several quantum perceptrons is a QNN. We show that a quantum perceptron with multi-qubit interactions  can do an XOR gate and prime number search from $3$ to $5$ bits, improving the performance of approximation power compared to the classical counterpart, as well as compared to standard QNNs (i.e. QNNs that include  nested perceptrons without multi-qubit interactions). Then, we show that quantum gates such as CNOT, Toffoli, and Fredkin can be implemented by QNN involving quantum perceptrons with multi-qubit interactions, thus reducing the circuit depth as it is not needed to add hidden layers.

\section*{Results}

\subsection*{Quantum perceptrons with multi-qubit potentials}
A quantum perceptron, or a quantum neuron, is the basic building block of a QNN. It can be constructed as a qubit that presents a nonlinear response to an input potential $\hat x_j$ in the excitation probability. This can be written as the following quantum gate acting on a $j$th qubit that encodes the quantum perceptron~\cite{Torrontegui19, Ban20}:
\begin{eqnarray}
 \label{gate}
 \hat{U}_j(\hat{x}_j; f) |0_j\rangle = \sqrt{1-f(\hat{x}_j)} |0_j\rangle + \sqrt{f(\hat{x}_j)} |1_j\rangle,
\end{eqnarray}
where  
\begin{eqnarray}
\label{sigmoid}
f(x) = \frac{1}{2} \left(1+\frac{x}{\sqrt{1+x^2}}\right).
\end{eqnarray}
corresponds to a nonlinear function.
The transformation in Eq.~(\ref{gate}) can be engineered, e.g., by evolving adiabatically the qubit with the Hamiltonian 
\begin{equation}\label{H}
\hat{H}  = \frac{1}{2}\left[\hat{x}_j  \hat{\sigma}^z_j + \Omega(t)\hat\sigma_j^x\right]
\end{equation} 
where $\hat{x}_j$ is the potential exerted by other neurons on the perceptron, and the applied external field $\Omega(t)$ leads to a tunable energy gap in the dressed-state qubit basis $|\pm\rangle$, with $\hat \sigma^x_j|\pm\rangle = \pm |\pm\rangle$. Typically,  
$\hat{x}_j = \sum_{i=1}^k (w_{ji} \hat{\sigma}^z_i) + b_j$~\cite{Torrontegui19, Ban20} which implies that the $j$th perceptron is coupled to a number $k$ of neurons (labelled with $i$) in the previous/input layer via standard spin-spin interactions. The Hamiltonian in Eq.~(\ref{H}) has the following reduced eigenstate (i.e. when the degrees of freedom of any other neuron are traced-out):
\begin{eqnarray}
\label{eigenstate}
|\Phi(x_j/ \Omega(t)) \rangle= \sqrt{1- f(x_j / \Omega(t))} |0_j\rangle + \sqrt{f(x_j / \Omega(t))} |1_j\rangle,~~~~
\end{eqnarray}
with $f(x)$, the excitation probability, in the form of Eq. (\ref{sigmoid}).
Specifically, to accomplish $\hat{U}_j(\hat{x}_j;f) |0_j\rangle$, one can use a Hadamard gate to firstly get the transformation $|0_j\rangle \rightarrow |+_j\rangle$ and  finally obtain $|\Psi\rangle = |\Phi(x_j/ \Omega(t_f)) \rangle$ at a certain time $t_f$ by evolving the system adiabatically with Hamiltonian~(\ref{H}). With the fixed trajectory always along with the instantaneous eigenstate of the Hamiltonian, one can deduce the external driving $\Omega(t)$ from the fast quasi-adiabatic passage~\cite{Torrontegui19}.
In this manner, the nonlinear activation function of the quantum perceptron is encoded in the probability of the excited state $P_j = f(x_j) = \frac{1}{2} (1+\langle \hat{\sigma}_j^z \rangle)$ during the adiabatic evolution~\cite{Torrontegui19}. 
In order to speed up the operation of this perceptron, one can also use inverse engineering techniques which directly impose conditions in the wave function evolution at the initial and final time instants, resulting in nonlinear response in the quantum perceptron~\cite{Ban20}. Correspondingly, a smoother control $\Omega(t)$ easily to be implemented experimentally can also be found. In addition, this accelerated activation mechanism by inverse engineering for the quantum neurons would reduce the decoherence and the variation of the input potential induced by neurons in the previous layer leading to enhanced performance. 

Now, we introduce a different type of potentials which rely on the possibility to implement multi-qubit interactions. In particular, we consider a potential of the kind
\begin{equation}\label{multipotential}
\hat{x}_j = \sum_{i=1}^k (w_{ji} \hat{\sigma}^z_i) + w_{\rm{m}} \hat\sigma_{l_1}^z ...  \ \hat\sigma_{l_n}^z + b_j.
\end{equation}
where $w_{\rm{m}}$ is a multi-qubit coefficient marked by the subscript $\rm{m}$, and $l_p \in [1, 2, ..., k]$ (namely, the term involving several Pauli matrices includes products of an arbitrary number of neurons in the previous input). For the sake of simplicity in the presentation, there is only a single multi-qubit term in Eq.~(\ref{multipotential}). However, this can include several products of distinct neurons in the input layer (i.e. additional multi-qubit terms). Later we will provide specific examples of these interactions associated to definite problems. 

In the following, we show that the multi-qubit potential enables tasks such as (i) Constructing XOR gates at the perceptron level (ii) Searching prime numbers and (iii) Encoding quantum gates. All these are implemented without hidden layers and/or ancillary qubits, thus, showing the significant role of multi-qubit potentials in the simplification of QNNs. 

\subsection*{XOR gate}
As a classical perceptron is a linear separator, a nonlinear logic gate such as the well-known XOR problem (i.e., the exclusive OR boolean function) requires at least one hidden layer to be implemented in classical neural networks \cite{perceptrons}. Now we show that a quantum perceptron with multi-qubit interactions in the neural potential is a nonlinear classifier. In particular, we illustrate the construction of an XOR gate by a single quantum perceptron with multi-qubit interactions. We also show that the lack of hidden layers prevents classical neural networks and standard QNNs to achieve the same task.

\begin{figure}[t]
	\begin{center}
		\scalebox{0.45}[0.45]{\includegraphics{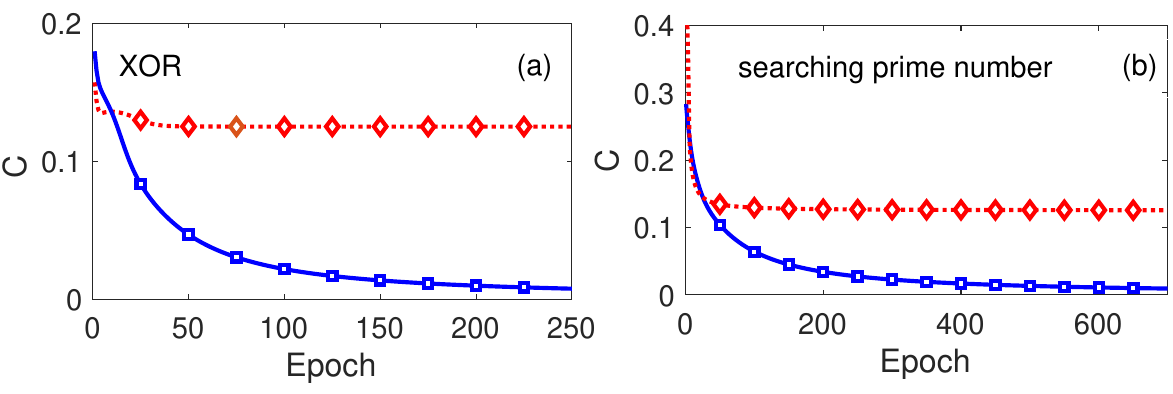}}
		\caption{\label{classical-tasks} In order to (a) encode an XOR gate and to (b) search prime numbers among the integers from $0$ to $7$, we show the value of the cost function at different epoches, using a multi-qubit interaction perceptron (solid blue) and a classical perceptron (dotted red). The latter is equivalent to a quantum perceptron with only two-qubit terms, which is shown from the comparison of the weights and bias during the training process, see Eqs.~(\ref{weights-classical}) - (\ref{partial-part}), where the learning rate is $\eta = 1.5$. The transfer function in Eq. (\ref{sigmoid}) is used for both cases. In the quantum perceptron, this function indicates that the system evolves adiabatically.
		}
	\end{center}
\end{figure}

As the values of the output neuron $0$ and $1$ cannot be separated linearly, a classical perceptron with two inputs and one output (i.e. without hidden layers) fails to solve the XOR gate. To show this, we use the standard gradient descent algorithm to train this simple classical perceptron with a sigmoidal activation potential $f(x)$, see Eq. (\ref{sigmoid}), and $x_j=\sum_{i=i}^kw_{ij}s_i+b_j$ with the classical input $s_i \in \{0,1\}$ whose cost function is in the form of the mean square value 
\begin{eqnarray}
\label{cost-function1}
C=\frac{1}{2N}\sum_{n=1}^N (y^{(n)} -t^{(n)})^2.
\end{eqnarray}
However, besides a sigmoid function, neurons in other forms of nonlinear activation functions, such as a ReLu function, can also create a neural network to learn the XOR gate. 
Here, $N=4$ determines the four possible examples  $00,10,01,11$, while $y^{(n)}$ and $t^{(n)}$ are the output and the target respectively, for the $n$th example. For simplicity, the subscript $j$ labelling one perceptron will be neglected in the following text. During the training, the parameters of the classical perceptron are updated after each epoch as 
\begin{eqnarray}
\label{weights-classical}
\nonumber
\tilde{w}_i &=& w_i - \eta \frac{\partial C}{\partial w_i} =  w_i - \frac{\eta}{N} \sum_{n=1}^N \left(y^{(n)} - t^{(n)}\right)  f'(x) s_i,
\\
\tilde{b} &=& b -\eta \frac{\partial C}{\partial b} = b - \frac{\eta}{N} \sum_{i=1}^N (y^{(n)} - t^{(n)}) f'(x),
\end{eqnarray} 
with the learning rate $\eta$. Any other nonlinear function $g(x)$, similar to $f(x)$, can be applied to train this perceptron in order to obtain the same approximation power. As it is shown in Fig. \ref{classical-tasks} (a), the cost function value of this classical perceptron stucks in $C=0.125$ (dotted-red line with superimposed diamonds for a better identification). 
As a classical perceptron can only converge on linearly separable data, it is not able to imitate the XOR function. In order to complete an XOR, a hidden layer with two neurons is needed. These two neurons can be regarded to perform an OR and a NAND gate.  

To implement an XOR gate with a quantum perceptron without the multi-qubit term in Eq.~(\ref{multipotential}), this is by using  the potential $\hat{x}= w_1 \hat{\sigma}_1^z + w_2 \hat{\sigma}_2^z +b $, one can follow the procedure described in Ref. \cite{Torrontegui19}. To encode an XOR gate, the neural potential is derived from the four basis states $|00\rangle$,  $|01\rangle$,  $|10\rangle$,  $|11\rangle$, which play the role of the four examples of the input in the XOR gate (namely, $00,10,01,11$).
In this case, as the input values of the XOR gate are bits, one can transform them into the measurement value of $\hat\sigma^z$ of the input qubits for the perceptron, i.e. $\sigma^{z}_{\rm{in}} = \frac{1}{2} (1+\langle\hat\sigma^z\rangle)$. Therefore, the input values $0$ and $1$ refer to the input states as the ground state $|0\rangle$ ($\langle\hat\sigma^z\rangle = -1$) and the excited state $|1\rangle$ ($\langle\hat\sigma^z \rangle=1$), respectively. 
Such a quantum perceptron without hidden layers and with only two-qubit interactions is equivalent to a classical perceptron, from the point of view of the following training process. 
Aiming at training it with the gradient descent method, one has to update the weights and bias as 
\begin{eqnarray}
\label{weights-quantum}
\tilde{w}_i &=& 
w_i - \frac{\eta}{N} \sum_{n=1}^N \left(y^{(n)} - t^{(n)}\right)  \frac{\partial y^{(n)}}{\partial x}  \frac{\partial x}{\partial w_i},
\nonumber
\\
\tilde{b} &=& b - \frac{\eta}{N} \sum_{i=1}^N \left(y^{(n)} - t^{(n)}\right) \frac{\partial  y^{(n)}}{\partial x},
\end{eqnarray}
where 
\begin{eqnarray}
\label{partial-part}
\nonumber
\frac{\partial  y^{(n)}}{\partial x} &=& \frac{1}{2} \left(\left\langle \frac{\partial \Psi}{\partial x} \bigl| \hat\sigma_z \bigr| \Psi \right\rangle + \left\langle \Psi \bigl | \hat\sigma_z \bigr | \frac{\partial \Psi}{\partial x} \right\rangle \right)
\\
&=& f'(x) 
\end{eqnarray}
and $|\Psi\rangle = \hat{U}(\hat{x},f) |0\rangle$  is the solution to the Schr\"{o}dinger equation driven by the  Hamiltonian~(\ref{H}) with $\hat{x}= w_1  \hat{\sigma}_1^z + w_2 \hat{\sigma}_2^z  +b $. In the previous equations we can see that  weights and bias are obtained in the same way as their classical counterparts. For that, one has to compare Eq.~(\ref{weights-classical}) with  Eqs.~(\ref{weights-quantum}) and  (\ref{partial-part}) provided that $\frac{\partial x}{\partial w_i}=\hat\sigma_i^z \rightarrow s_i$. This indicates that  a single quantum perceptron with two-qubit interactions and the basis states $|00\rangle$,  $|01\rangle$,  $|10\rangle$,  $|11\rangle$  as input is equivalent to a classical perceptron.

In order to implement the XOR gate with this quantum perceptron, one would need to perform two adiabatic passages with different controls $\Omega(t)$, as well as to use different neural potentials in each passage by appropriately changing the weights and biases~\cite{Torrontegui19}. Equivalently, one could also do the XOR gate by including one hidden layer with two additional quantum neurons and the application of a single $\Omega(t)$. It is worth mentioning that Eq. (\ref{partial-part}) holds only for a quantum perceptron instead of a QNN with hidden layers. If a QNN has more layers, one only needs to measure the output qubit value  
\begin{eqnarray}
\label{output}
y=P(x_{\rm{out}}) = \frac{1}{2} (1+\langle\Psi| \hat\sigma^z_{\rm{out}}|\Psi\rangle)
\end{eqnarray}
instead of the intermediate ones, i.e., $|\Psi\rangle = \hat{U}_{\textrm{tot}} |0\rangle$ 
with $\hat{U}_{\textrm{tot}} = \Pi_{j=1}^M \hat{U}_j$ where $M$ is the total number of quantum perceptrons in a QNN. Otherwise, shot noise is introduced by measuring the neurons in the hidden layer. 

This procedure to do the XOR gate can be simplified when considering a quantum perceptron with multi-qubit interactions. We can find the output of the quantum perceptron by using Eq. (\ref{output}),  where $|\Psi\rangle = \hat{U}(\hat{x};f) |0\rangle$. In the unitary transformation implemented by the quantum perceptron gate in the Heisenberg picture
\begin{eqnarray}
\label{output-unitary}
\hat{U}^\dag \hat{\sigma}^z \hat{U} = [1-2f(\hat{x})] \hat{\sigma}^z + 2 \sqrt{f(\hat{x})[1-f(\hat{x})]} \hat{\sigma}^x,
\end{eqnarray}
leads to $y = P(x) = f(x)$, where $y=1$, if $x>1/2$; and $y=0$, if $x \leq 1/2$.  In particular, we explore the following multi-qubit potential
\begin{equation}\label{potential-XOR}
\hat{x}= w_1  \hat{\sigma}_1^z + w_2 \hat{\sigma}_2^z  +b  + w_{\rm{m}} \hat{\sigma}_1^z \hat{\sigma}_2^z.
\end{equation} 
In this case, the weights $w_1$, $w_2$ and $b$ are updated as in Eqs.~(\ref{weights-quantum}), while the updating formula for the weight of the $w_{\rm{m}} \hat{\sigma}_1^z \hat{\sigma}_2^z$ term is
\begin{eqnarray}
\label{weights-multi-qubit}
\tilde{w}_{\rm{m}} &=& w_{\rm{m}}- \frac{\eta}{N} \sum_{n=1}^N \left(y^{(n)} - t^{(n)}\right)  \frac{\partial y^{(n)}}{\partial x}  \hat{\sigma}^z_1 \hat{\sigma}^z_2.
\end{eqnarray}
Considering the construction of an XOR gate from the equations 
\begin{eqnarray}
\label{weights-inputs}
&&0\times w_1 + 0 \times w_2 + b \leq \frac{1}{2} \Leftrightarrow b -  \frac{1}{2} \leq 0, 
 \label{condition1}
\\
&&0\times w_1 + 1 \times w_2 + b > \frac{1}{2} \Leftrightarrow b -  \frac{1}{2} >  - w_2,
 \label{condition2}
\\
&&1\times w_1 + 0 \times w_2 + b > \frac{1}{2}  \Leftrightarrow b -  \frac{1}{2} > - w_1,
 \label{condition3}
\\
&&1\times w_1 + 1 \times w_2 + b + w_{\rm{m}} \leq \frac{1}{2} \Leftrightarrow b -  \frac{1}{2}  + w_{\rm{m}} \leq  -w_1 - w_2,
 \label{condition4}
\end{eqnarray}
we find that Eq. (\ref{condition2}) - (\ref{condition4}) are contradictory, if the mutli-qubit interaction term does not exist. However, one can always find approprint values for $w_{\rm{m}}$ to satisfy the above inequalities. The existence of $w_{\rm{m}} \hat{\sigma}_1^z \hat{\sigma}_2^z$ in the neural potential enables the quantum perceptron to construct an XOR gate as a nonlinear separator. 
We test the cost function value (see, Eq.~(\ref{cost-function1})) for our quantum perceptron and find that $C< 1\%$ occurs at epoch $=197$ as shown in Fig.~\ref{classical-tasks}~(a) (solid-blue line with squares). From the numerical calcluation, we can see that the cost function value continues to decrease. On the contrary the classical and qubit-qubit interaction perceptron are not able to produce an XOR gate, see plateau behavior of the cost function (dotted-red) that remains constant after the epoch $\sim 50$.

\subsection*{Searching prime numbers}
\begin{figure}[t]
	\begin{center}
		\scalebox{0.20}[0.20]{\includegraphics{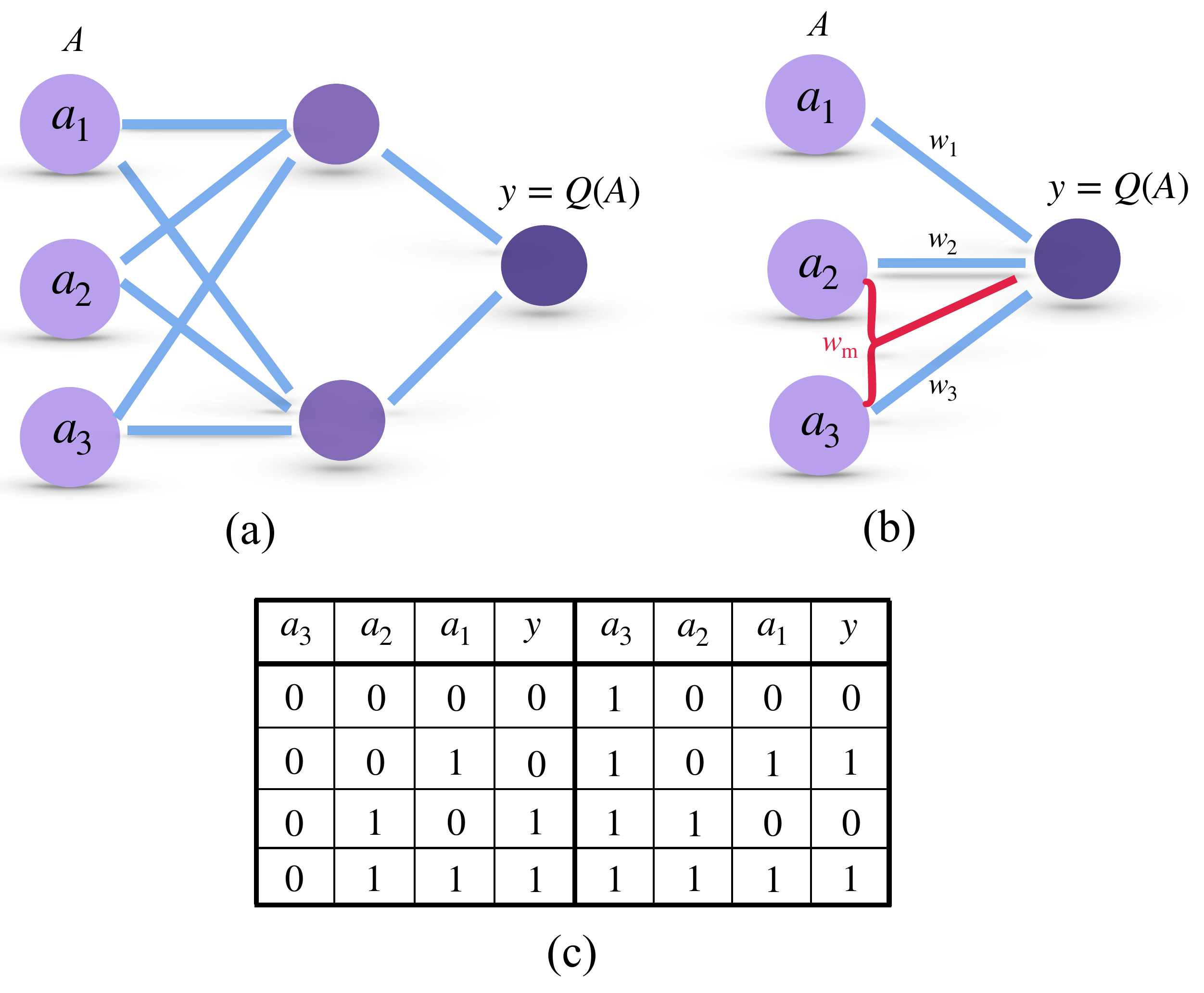}}
		\caption{\label{schematic-prime}	(a) Schematic configuration of a classical neural network with one hidden layer of two neurons for the task to search prime numbers of the input number $A = (a_1, a_2, a_3)\in \{0, 1, ..., 7\}$ with $3$ bits. After well trained, the network gives the output $y=Q(A)$.
(b) The same task can be achieved by a QNN which has multi-qubit interactions in the neural potential (red junction involving second and third neurons) without hidden layers. (c) Truth table for prime-number search with $3$ input bits for a QNN. The input values $A = (a_1, a_2, a_3)$ are the binary numbers for the integers from $0$ to $7$. The output value $y = Q(A)$ is $1$ for prime numbers and $0$ for non-prime ones.}
	\end{center}
\end{figure}

In Ref.~\cite{Torrontegui19}, a specific example of two, three, and four perceptrons per layer was illustrated to detect prime numbers from $0$ to $2^i -1$ where $i$ ranges from $i=3$ to $i=7$ bits. In particular, it was shown that a QNN with two perceptrons --one in the hidden layer and the other one as the output-- can classify prime numbers from $0$ to $7$ (3 bits). This exemplifies the better performance of QNNs compared with classical ones where a hidden layer with two neurons are necessary to accomplish the same task. A scheme of these networks with $3$ bits in input is shown figure in Fig. \ref{schematic-prime} (a) and (b), while details regarding QNN-training to search prime numbers are in Methods.

Now, we introduce the multi-qubit term into the neural potential and find that the same task is achieved by a QNN without hidden layers, i.e. at the single quantum perceptron level. To search a prime number for $3$ bits (Truth table listed in Fig. \ref{schematic-prime} (c)) using a single quantum perceptron we consider the potential
\begin{equation}
\label{potential-primesearch-3bits}
\hat{x}= w_1  \hat{\sigma}_1^z + w_2 \hat{\sigma}_2^z + w_3 \hat{\sigma}_3^z +b  + w_\textrm{m} \hat{\sigma}_2^z \hat{\sigma}_3^z.
\end{equation} 
As we demonstrate later, adding the multi-qubit term $w_\textrm{m} \hat{\sigma}_2^z \hat{\sigma}_3^z$ is enough to fulfill the prime numbers searching task. In this respect, one can include additional multi-qubit terms in the neural potential. However, we use the simplest example in Eq.~(\ref{potential-primesearch-3bits}) to simplify the network and training process.

We note that when the input of a quantum perceptron that presents only two-qubit interactions are the basis states, this is equivalent to the classical perceptron as shown in Eqs.~(\ref{weights-classical}) - (\ref{partial-part}). For this reason, now we compare the cost function of a classical perceptron with the cost function of a quantum perceptron using the potential in Eq.~(\ref{potential-primesearch-3bits}) for the specific purpose of searching prime numbers from $0$ to $7$. As it is shown in Fig.~\ref{classical-tasks} (b) the quantum perceptron  achieves $C < 1\%$ at epoch$=667$ (solid-blue with squares), while the $C$ value saturates to $0.126$ in the classical counterpart (dotted-red with diamonds). This indicates that our perceptron achieves the searching task without hidden layers. We have also verified the ability of the quantum perceptron by minimizing the cost function to an acceptable error for searching prime numbers for $4$ bits with the  potential 
\begin{equation}
\label{potential-primesearch-4bits}
\hat{x}= w_1 \hat{ \sigma}_1^z + w_2 \hat{\sigma}_2^z + w_3 \hat{\sigma}_3^z + w_4 \hat{\sigma}_4^z + b  + w_\textrm{m} \hat{\sigma}_2^z \hat\sigma_3^z,
\end{equation} 
and for $5$ bits with 
\begin{eqnarray}
\label{potential-primesearch-5bits}
\hat{x}&=& w_1 \hat{ \sigma}_1^z + w_2 \hat{\sigma}_2^z + w_3 \hat{\sigma}_3^z + w_4 \hat{\sigma}_4^z + w_5 \hat{\sigma}_5^z + b  
\nonumber
\\
&&+ w_\textrm{m} \hat{\sigma}_2^z \hat{\sigma}_3^z .
\end{eqnarray} 
The multi-qubit terms in the above two neural potentials for $4$ and $5$ bits are chosen due to the fact that the cost function result in $C<1\%$ at epoch $=1692$ for Eq.~(\ref{potential-primesearch-4bits}) and epoch $=1698$ for Eq.~(\ref{potential-primesearch-5bits}) with the consideration of adopting the minimal number of multi-qubit terms.
Further training the above quantum perceptrons presents that the value of $C$  continue to decrease, proving the success to achieve these tasks. 

\subsection*{Quantum gates}
Now we show that one can use a QNN with multi-qubit interactions to construct quantum gates such as CNOT, Toffoli, and Fredkin gates without the necessity of hidden layers. In comparison, one can demonstrate that a QNN with only two-qubit interactions cannot construct the above mentioned quantum gates in the same conditions. This will be shown later. As quantum gates are reversible, the number of input and output qubits of quantum gates should be equal. Hence, the number of the neurons in the input equals that in the output.
Correspondingly, the cost function is changed into
\begin{eqnarray}
\label{cost-function2}
C=\frac{1}{2Nk}\sum_{n=1}^N \sum_{i=1}^k (y_i^{(n)} -t_i^{(n)})^2,
\end{eqnarray}
where $k$ is the number of quantum perceptrons. 

\begin{figure}[t]
	\begin{center}
		\scalebox{0.2}[0.2]{\includegraphics{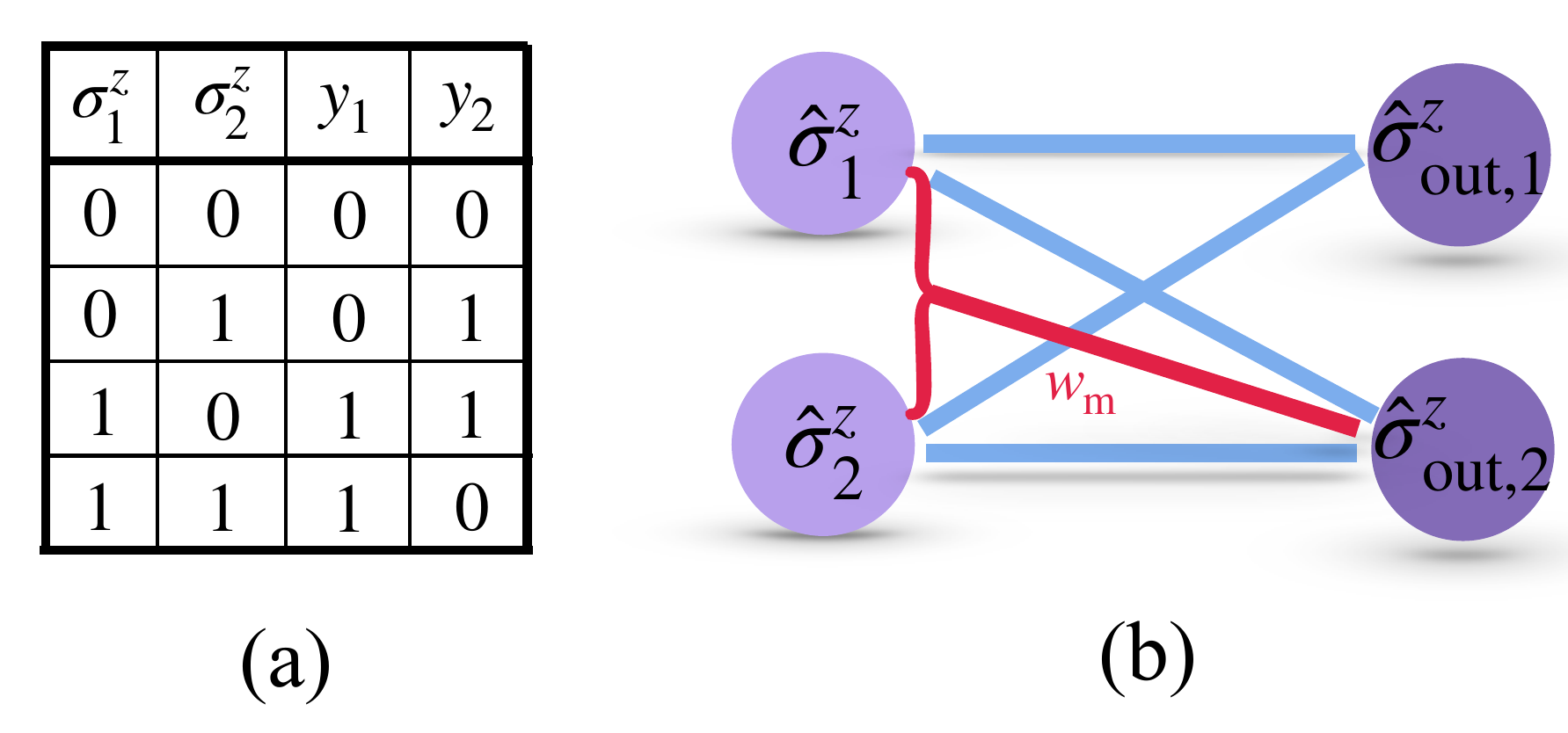}}
		\caption{\label{schematic-CNOT} Truth table of a CNOT gate (a) constructed by a QNN illustrated in a schematic configuration (b) with the multi-qubit interaction term $w_{\rm{m}} \hat\sigma_1^z \hat\sigma_2^z$.
		}
	\end{center}
\end{figure}

\begin{figure}[ht]
	\begin{center}
		\scalebox{0.45}[0.45]{\includegraphics{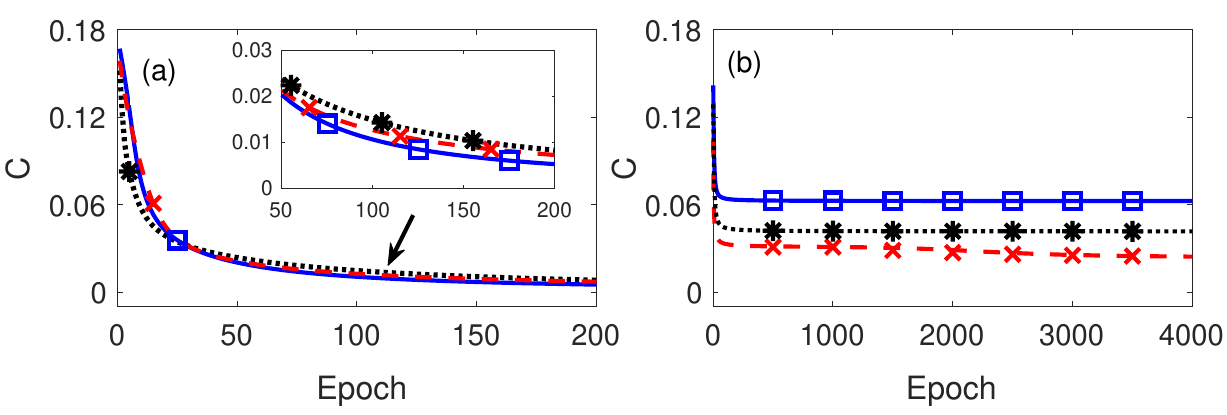}}
		\caption{\label{quantum-gates} Cost function at different epoch for (a) QNN with the multi-qubit term and (b) QNN with ony two-qubit interactions. In particular the solid-blue line is the cost function obtained to construct a CNOT gate, a Toffoli gate (dashed-red line) and a Fredkin gate (dotted-black line). Illustrated in the inset of (a), the cost function values reach $C<1\%$ at  $116$th, $140$th, $162$th epoch for the above three quantum gates, respectively, indicating that these gates can be constructed by QNNs with multi-qubit interaction without hidden layers. In contrast, $C$ values go to their respective plateau by quantum perceptrons with two-qubit interactions.}
		
	\end{center}
\end{figure}

For CNOT gate where $k=2$, the truth table and the schematic configuration of the QNN is shown in Fig. \ref{schematic-CNOT} (a) and (b) respectively, where each perceptron (output neuron) gives the output value  $y_{i} = \frac{1}{2}(1+\langle\hat\sigma_{\rm{out,i}}^z\rangle)$.  
The first perceptron should have the same value as the first input neuron, while the second one aims to achieve the same value as the XOR output of two input neurons, i.e. $\hat\sigma^z_{\rm{out,2}} = \hat\sigma_1^z \oplus \hat\sigma_2^z$ \cite{quantum-computation-book}. The neural potential of the second perceptron is 
\begin{eqnarray}
\label{H-CNOT}
\hat{x}_2 = w_{21} \hat\sigma_1^z  + w_{22} \hat\sigma_2^z  + b_2 + w_{\rm{m}} \hat\sigma_1^z\hat\sigma_2^z, 
\end{eqnarray} 
and the first perceptron has two-qubit interaction terms. We find that this is the most concise choice, as it requires least connectivity.
We obtain the cost function value (using Eq.~(\ref{cost-function2}) $C<1\%$ at epoch $=116$ and $C\rightarrow 0$ at a larger epoch, as shown in Fig. \ref{quantum-gates} (a) (solid-blue with squares) indicating the satisfaction to train a CNOT gate well. On the contrary, training such a CNOT gate by a QNN with only two-qubit interaction terms leads to a $C$ tending to $0.0625$ at large epoch, see Fig.~\ref{quantum-gates} (b). This indicates that a standard QNN cannot encode the CNOT without hidden layers.

\begin{figure}[b]
	\begin{center}
		\scalebox{0.20}[0.20]{\includegraphics{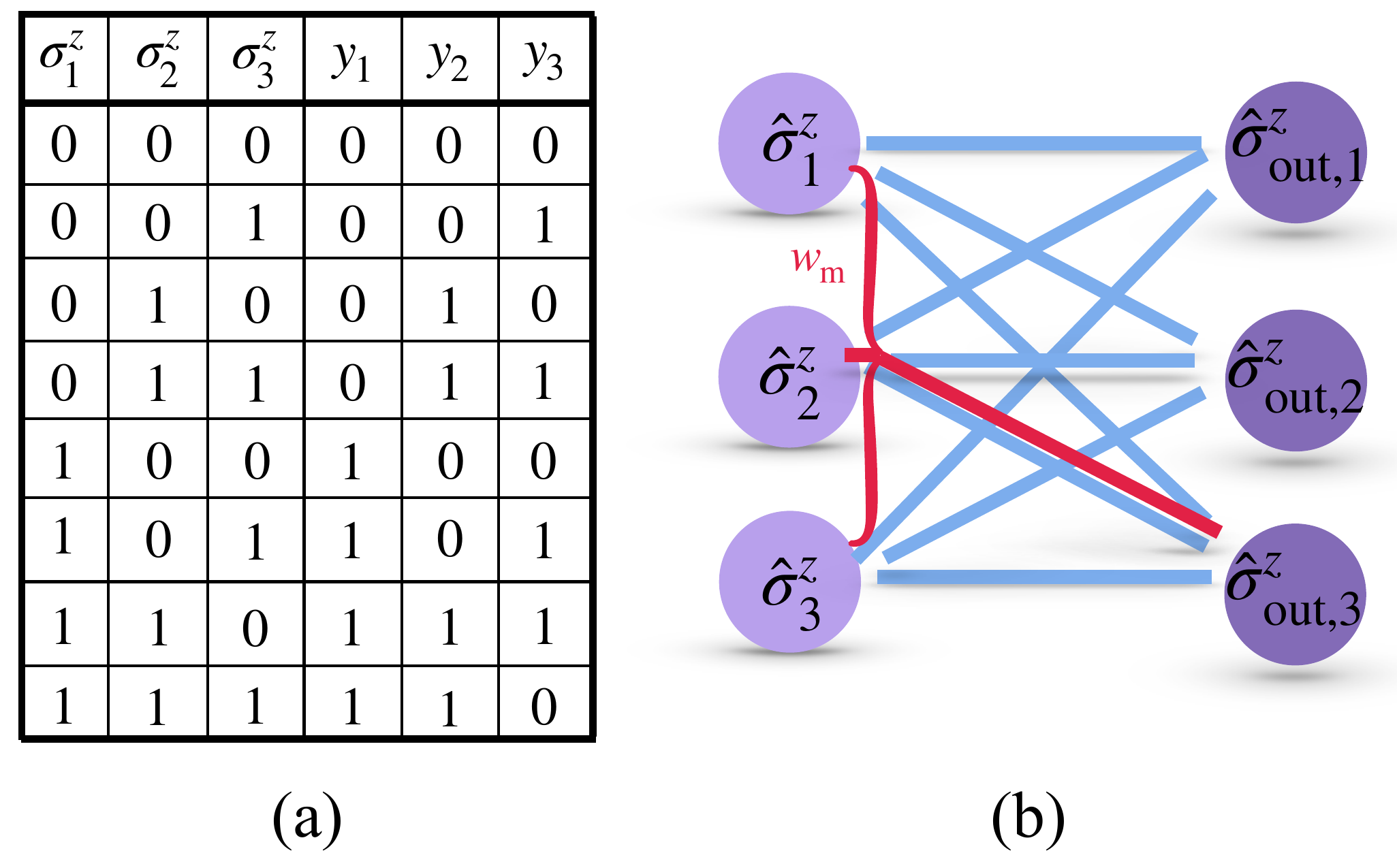}}
		\caption{\label{schematic-toffoli} Truth table of a Toffoli gate (a) constructed by a QNN illustrated in a schematic configuration (b) with the multi-qubit interaction term $w_{\rm{m}} \hat\sigma_1^z\hat\sigma_2^z \hat\sigma_3^z$.
		}
	\end{center}
\end{figure}
A QNN with multi-qubit terms also works for constructing a Toffoli gate without hidden layers. Truth table, and QNN for    Toffoli are shown in Fig. \ref{schematic-toffoli} (a) and (b). Known as Controlled-Controlled-Not gate, a Toffoli gate has 3-qubit inputs and outputs. In the QNN, the outputs of the first two qubits should be the same value as their inputs $\hat\sigma_{\rm{out,i}}^z = \hat\sigma^z_i$ ($i = 1, 2$), while the output of the third qubit aims at $\hat\sigma_{\rm{out,3}}^z = \hat\sigma_3^z \oplus \hat\sigma^z_1 \hat\sigma^z_2$ \cite{quantum-computation-book}. 
In this case, we propose the following neural potential of the third perceptron
\begin{eqnarray}
\label{H-toffoli}
\hat{x}_3 &=& w_{31} \hat\sigma_1^z  + w_{32} \hat\sigma_2^z  + w_{33} \hat\sigma_3^z 
\nonumber
\\&&+ b_3
+ w_\textrm{m} \hat\sigma_1^z \hat\sigma_2^z \hat\sigma_3^z,
\end{eqnarray} 
with a minimal number of multi-qubit terms.
Demonstrated in Fig. \ref{quantum-gates} (a) and (b), the value $C<1\%$ (dashed-red line with crosses) occurs at epoch $=140$ and goes to $0$ at large epoch during the training of a Toffoli gate. In close similarity with the previous case, a standard QNN without hidden layers fails to achieve the Toffoli gate, as $C$ saturates at $0.0238$.

\begin{figure}[t]
	\begin{center}
		\scalebox{0.2}[0.2]{\includegraphics{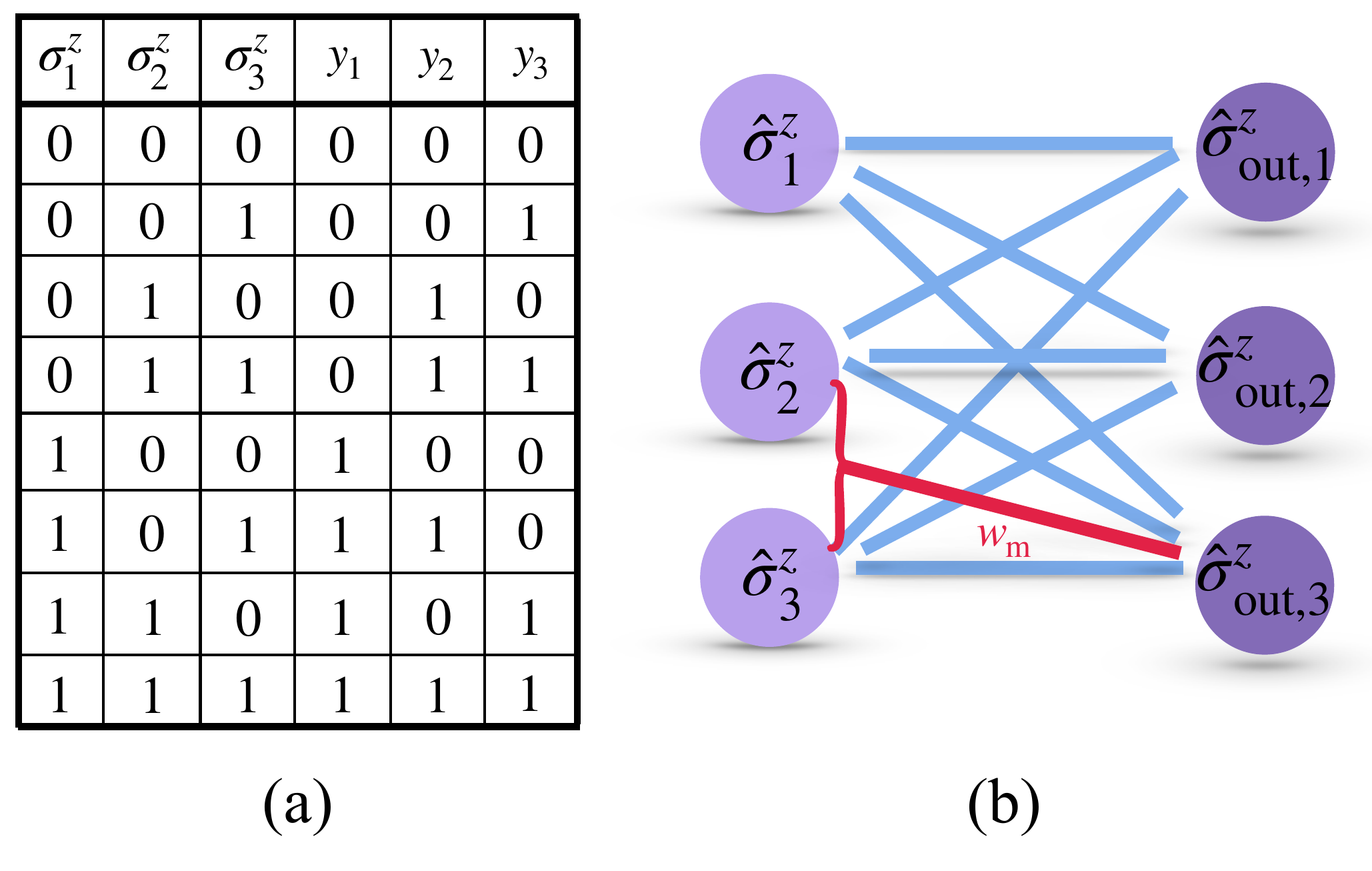}}
		\caption{\label{schematic-fredkin}  Truth table of a Fredkin gate (a) constructed by a QNN illustrated in a schematic configuration (b) with the multi-qubit interaction term $w_{\rm{m}} \hat\sigma_2^z\hat\sigma_3^z$.
		}
	\end{center}
\end{figure}

Another example that can be successfully constructed by our QNNs including multi-qubit interactions without hidden layers is the Fredkin gate (also known as Controlled-SWAP gate) whose truth table and schematic configuration are in Fig.~\ref{schematic-fredkin} (a) and (b).
In this case, the adopted  potential of the third perceptron is 
\begin{eqnarray}
\label{H-fredkin}
\hat{x}_3 &=& w_{31} \hat\sigma_1^z  + w_{32} \hat\sigma_2^z  + w_{33} \hat\sigma_3^z 
\\
&&+ b_3 + w_\textrm{m} \hat\sigma_2^z \hat\sigma_3^z,
\end{eqnarray} 
According to the numerical calculation, the cost function value $C$ tends to zero, shown in Fig. \ref{quantum-gates} (a) (dotted-black line with asterisks), indicating that these gates can be constructed by quantum perceptrons without hidden layers. Again, $C$ saturates at the value $0.0417$ at large epoch for a standard QNN without hidden layers, see Fig. \ref{quantum-gates} (b).

\section*{Discussions}
We have shown different applications of QNNs which possess multi-qubit interactions.  Meanwhile, the performance is enhanced compared to the same topology of a QNN without multi-qubit interactions. The multi-qubit interaction terms induce connectivities among quantum perceptrons that deviate from the current network paradigm of additive activations. It is due to the multi-qubit terms in the potentials that one can avoid the presence of some hidden layer without sacrificing approximative power. Such architecture allows us to address the connectivity challenge in scaling up QNNs. Moreover, the simple configuration helps to control the efficiency of the training processes. During the training process of all the examples shown above, the activation function (Eq. \ref{sigmoid}) based on the adiabatic evolution of the system is used. Instead, one may use {\it shortcuts to adiabaticity} to accelerate the formation of the activation function in physical registers~\cite{Ban20}.

The Hamiltonian of the $j$th perceptron given by Eqs.~(\ref{H}) where the neural potential expressed by Eq. ~(\ref{multipotential}) corresponds to the linear Ising Hamiltonian with higher order interactions. Such a model is present in distinct quantum platforms \cite{Donskaya1984,Kumar20,Hartmann17,Hartmann22,Lanyon11,Borjans20} experimentally and theoretically, although the developments in different platforms vary. An optical Ising machine hosting adjustable four-body interaction with all-to-all connections over a large number of spins was experimentally demonstrated in Ref. \cite{Kumar20}. Moreover, the four-body interactions are designed to be realized via superconducting quantum interference devices (SQUIDs) \cite{Hartmann17}, while a single shot method for executing an i-Toffoli gate which is a three-qubit gate with two control and one target qubits was proposed in Ref.  \cite{Hartmann22},  with the application of currently existing superconducting hardware. 
Using resonant microwave-mediated interactions between distant electron spins to implement multi-qubit potentials \cite{Borjans20} marks an important milestone for all-to-all qubit connectivity and scalability in silicon-based quantum circuits. Quantum evolutions governed by terms involving up to six-qubit interactions has already been implemented in trapped-ion systems~\cite{Lanyon11} where the value of qubit-qubit interactions can be tuned, allowing the same architecture to be used to implement different types of gates and leading to a variety of quantum gates in the system. Each of the examples considered above corresponds to a particular case of multi-qubit interaction and its physical implementation would depend on the specific considered platform. The development in the quantum hardwares highlights the potential for practical implementations of multi-qubit potentials in QNNs. The promotion of quantum hardware and QNN protocols in respective fields will boost mutual developments.

\section*{Methods}
\subsection*{Hamiltonian and training of the neural networks}
In the article, we propose a QNN with multi-qubit interactions where the Hamiltonian Eq. (\ref{H}) of one perceptron has the neural potential expressed in Eq. (\ref{multipotential}). 
In preparation for a QNN to demonstrate advantages in comparison with classical counterparts, we need to develop concepts to improve scalability of QNNs where controlling the network depth becomes crucial. One main objective to develop a QNN is to therefore minimize the depth without sacrificing approximative power. To this end, we explore deviations from the current network paradigm of additive activations and include multi-qubit interactions in the neural potential leading to a reduction of the network depth. 
In all the calculations, we use the standard gradient descent to train the neural networks with the cost function Eq. (\ref{cost-function2}). Correspondingly, the weight of multi-qubit interactions term $w_{\rm{m}} \hat\sigma_{l_1}^z ...   \hat\sigma_{l_n}^z$ in the neural potential Eq. (\ref{multipotential}) should be updated as 
\begin{eqnarray}
\label{weights-multi-qubit-general}
\tilde{w}_{\rm{m}} &=& w_{\rm{m}}- \frac{\eta}{N} \sum_{n=1}^N \left(y^{(n)} - t^{(n)}\right)  \frac{\partial y^{(n)}}{\partial x}  \hat{\sigma}^z_{l_1} ... \hat{\sigma}^z_{l_n}.
\end{eqnarray}

\subsection*{Training for prime numbers search}
To train a QNN for the task of searching prime numbers, a set of $2^i$ pairs for $i$ bits containing the input and output values $\{A^{(n)}, y^{(n)}\}^{2^i}_{n=1}$ is taken, where the inputs are binary numbers $A^{(n)} = (a_1, a_2, ..., a_i)$ corresponding to the integers belonging to the set $\in\{0, 1, ...,2^i -1\}$. As the targets are $t \in \{0,1\}$, the output of the network $y^n = Q(A^{(n)}) = 1$, if and only if the input number is prime.

\section*{Data availability}
The datasets used and analysed during the current study available from the corresponding author on reasonable request.

{}

\section*{Acknowledgements}
We acknowledge financial support from Spanish Government via PGC2018-095113-B-I00 (MCIU/AEI/FEDER, UE), Basque Government via IT986-16, as well as from QMiCS (820505) and OpenSuperQ (820363) of the EU Flagship on Quantum Technologies, and the EU FET Open Grant Quromorphic (828826). 
Y. B. acknowledges the CDTI within the Misiones 2021 program and the Ministry of Science and Innovation under the Recovery, Transformation and Resilience Plan – Next Generation EU under the project “CUCO: Quantum Computing and its Application to Strategic Industries”.
E. T. acknowledges  the Ramón y Cajal (RYC2020-030060-I) research fellowship, the Spanish Government via the project PID2021-126694NA-C22 (MCIU/AEI/FEDER, EU), CSIC Research Platform PTI-001, CAM/FEDER Project No. S2018/TCS-4342 (QUITEMAD-CM), and  by Comunidad de Madrid-EPUC3M14. J.~C. acknowledges the Ram\'{o}n y Cajal   (RYC2018-025197-I) research fellowship, the financial support from Spanish Government via EUR2020-112117 and Nanoscale NMR and complex systems (PID2021-126694NB-C21) projects,  the ELKARTEK project Dispositivos en Tecnolog\'i{a}s Cu\'{a}nticas (KK-2022/00062), and the Basque Government grant IT1470-22.

\section*{Author contributions statement}
Y. Ban developed the theoretical formalism, performed the analytic calculations and the numerical simulations. E. Torrontegui verified the method. J. Casanova supervised the project. All the authors contributed to the final version of the manuscript.

\section*{Additional information}
\textbf{Competing interests}: The authors declare no competing financial interests.

\end{document}